**Living porous ceramics for bacteria-regulated gas sensing and carbon capture**

Alessandro Dutto, [1] Anton Kan, [1] Zoubeir Saraw, [1] Aline Maillard, [1] Daniel Zindel, [2] André R. Studart [1]

[1] Complex Materials, Department of Materials, ETH Zürich, 8093 Zürich, Switzerland

[2] Laboratory of Physical Chemistry, ETH Zürich, 8093 Zürich, Switzerland

**Abstract**

Microorganisms hosted in abiotic structures have led to engineered living materials that can grow, sense and adapt in ways that mimic biological systems. Although porous structures should favor colonization by microorganisms, they have not yet been exploited as abiotic scaffolds for the development of living materials. Here, we report porous ceramics that are colonized by bacteria to form an engineered living material with self-regulated and genetically programmable carbon capture and gas-sensing functionalities. The carbon capture capability is achieved using wild-type photosynthetic cyanobacteria, whereas the gas-sensing function is generated utilizing genetically engineered *E. coli*. Hierarchical porous clay is used as ceramic scaffold and evaluated in terms of bacterial growth, water uptake and mechanical properties. Using state-of-the-art chemical analysis techniques, we demonstrate the ability of the living porous ceramics to capture $CO_2$ directly from the air and to metabolically turn minute amounts of a toxic gas into a benign scent detectable by humans.

**Introduction**

Engineered living materials (ELMs) harness the biological activity of microorganisms to imbue synthetic structures with adaptive, sensing and decision-making functionalities. [1-5] Examples of ELMs include self-healing concrete, [6] living sensors for water quality control, [7] regenerative robotic skins, [8] and self-lubricating contact lenses. [9] The microorganisms responsible for these living properties vary from bacteria to algae and fungi, which can be harvested directly from the environment or genetically engineered for the function of interest. Engineered microorganisms can harness biological activity in a programmable manner to add diverse functionalities to ELMs, leading to applications in catalysis, energy conversion, sensing, electronics, and biomedicine. [2,10-20] Despite these remarkable examples, the potential of ELMs in generating and keeping living functionalities strongly depends on the host material. For the ELM to be functional and remain alive it is essential to provide a host material for the cells to proliferate and maintain their metabolic activity. While hydrogels and particle networks have been widely used as host structures, they may limit the growth of microorganisms and their access to oxygen and nutrient-containing media. This calls for novel scaffold materials that promote the colonization by microorganisms, supply of nutrients, and harness the resulting biological activity of the engineered living material.

Porous ceramics have been extensively used as scaffolds for the in-growth of cells for tissue engineering applications. [21,22] By tuning the porosity and pore size of the structure, it is possible to create scaffolds that promote cell proliferation while ensuring sufficient nutrient supply for their growth and biological activity. This is often achieved using scaffolds with a hierarchical porous architecture, in which small pores below 10 $\mu$m generate the high surface area required for cell adhesion [23] whereas large pores with 50-1000 $\mu$m provide the high permeability needed for vascularization and continuous supply of oxygen and nutrients. [24] A broad range of techniques has been utilized for the fabrication of hierarchical porous ceramics, including 3D printing, foaming, emulsion templating and freeze casting. [25-27] 3D printing of wet foams and emulsions has been shown to be a particularly suitable approach to independently control pore sizes at multiple length scales. [28] Using these hierarchical porous ceramics as host scaffolds for wild-type or engineered microorganisms should lead to mechanically robust living materials with high metabolic activity and tunable functionalities.

The functionalities of engineered living materials are primarily controlled by the activity of the microorganisms hosted within an abiotic structure. Depending on the function of interest, the desired activity might already be encoded in wild-type microorganisms or may require genetic modification to create new microbial strains. Carbon capture is an example of a functionality of ELMs that has been achieved using wild-type photosynthetic microorganisms, such as microalgae and cyanobacteria. [29-32] In this case, the natural metabolism of the microorganism is harnessed to convert $CO_2$ from the air into organic molecules that can be stored within the cells. Beyond native species, the range of functionalities of ELMs can be significantly extended by programming the hosted microorganisms with synthetic biology tools. [33,34] Illustrative examples are living sensors that contain bacteria engineered to detect, transduce, and amplify chemical signals using programmable genetic circuits. [14,35] Overall, the wide spectrum of microorganisms, synthetic biology tools and manufacturing processes currently available provide extensive opportunities to develop living materials engineered with functions unavailable to traditional materials.

Here, we develop and study living porous ceramics with gas-sensing and carbon-capture functionalities regulated by metabolically active bacteria. Inspired by the water transport mechanism of trees, the porous structure is designed to pump and distribute nutrient-rich water for the growth and metabolism of microorganisms embedded within the living material. First, we study the water transport, bacterial proliferation and mechanical stability of the porous ceramic structure. Next, ureolytic bacteria are used to mechanically reinforce the porous structure via a room-temperature biocementation process. The porous ceramics are then loaded with photosynthetic microorganisms and investigated in terms of carbon capture capabilities. Finally, engineered bacteria are implemented in the porous ceramics to create a living material that indicates the presence of a toxic gas by generating a benign smell detectable by the human nose.

**Results and discussion**

The envisioned living porous ceramics should display a large fraction of open, interconnected pores to enable growth of the microorganisms and continuous intake of nutrient-rich culture medium (Figure 1). Bacterial growth is facilitated by selecting ceramic particles featuring biocompatible and hydrophilic surface chemistry. To fulfill this criterion, naturally occurring clay was chosen to be the solid phase of the porous ceramic. Besides its environmental-friendly nature, clays are widely available and have been extensively used in the building industry. Moreover, several processing techniques exist for the processing of clay-based ceramics with controlled porosity and pore sizes (Figure 1a).

Ceramics with carbon capture and gas-sensing functionalities can be created by growing bacteria with the desired metabolic features within the porous structure (Figure 1b,c). This is possible using either wild-type or genetically modified microorganisms. As an essential natural process of the Earth's carbon cycle, the consumption of $CO_2$ during photosynthesis is a metabolic feature found in a broad range of autotropic wild-type microorganisms. For gas sensing, the biological activity of the microorganisms needs to be engineered for the detection, transduction, and amplification of the molecules of interest, which so far has mostly been demonstrated for chemicals dissolved in liquids. [14,36,37]

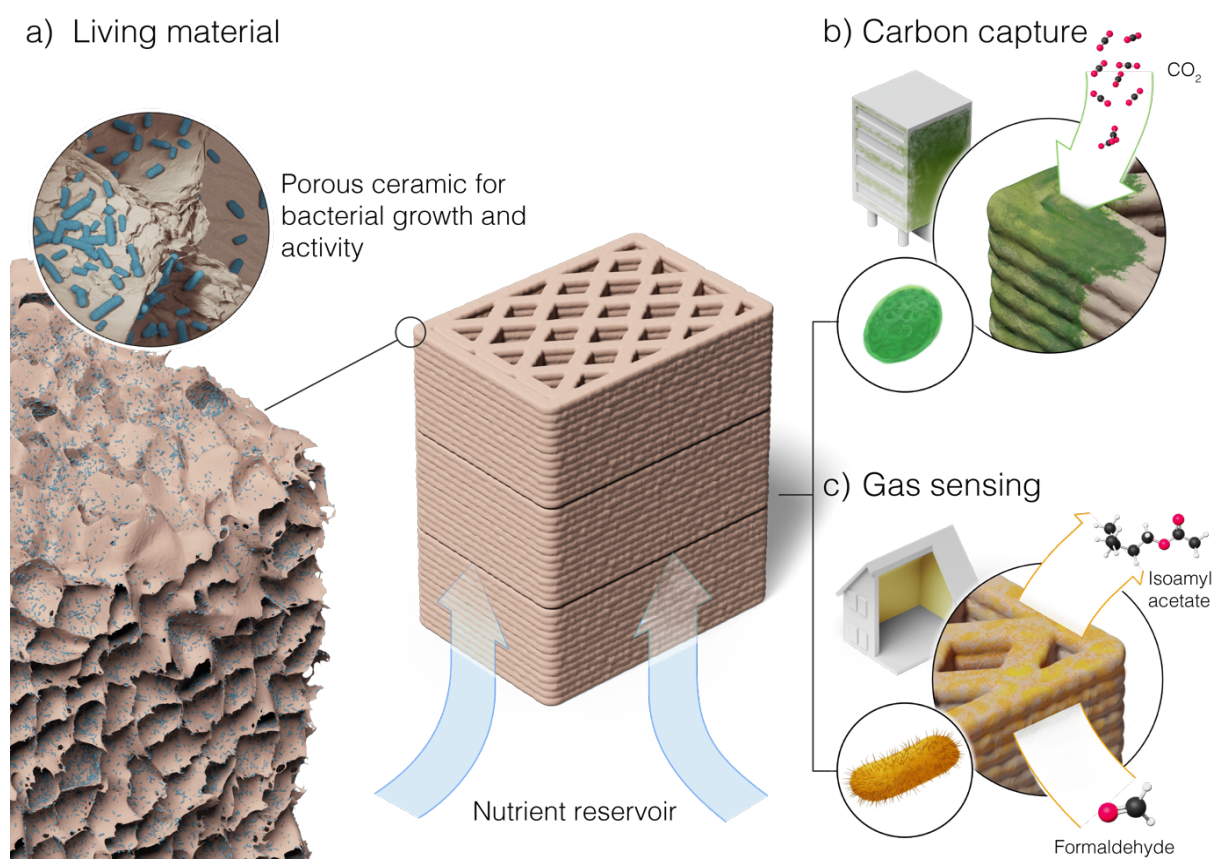

**Figure 1. Living porous ceramic for carbon capture and gas sensing.** The potential use of living ceramics as building materials is highlighted. (a) The porous ceramic serves as scaffold for the growth and activity of specific microorganisms. (b) Carbon capture and (c) gas-sensing capabilities are achieved by using wild-type photosynthetic cyanobacteria or engineered microorganisms designed as biosensors.

To create and study the proposed living porous ceramics, four microorganisms were selected: (i) an *Escherichia coli* strain engineered to produce green fluorescent protein (GFP), (ii) the wild-type ureolytic soil bacterium *Sporosarcina pasteurii*, (iii) a wild-type, cyanobacteria *Synechococcus sp.* (PCC 7002), and (iv) an *Escherichia coli* strain engineered for gas sensing (Figure 2a). The GFP-producing *E. coli* was used as a model microorganism to investigate the growth and colonization of the porous structure by bacteria. Wild-type *S. pasteurii* and *Synechococcus sp.* are two broadly available microorganisms that were used for mechanical reinforcement via biocementation and for carbon capture through photosynthesis, respectively. The gas-sensing *E. coli* strain was engineered to detect low concentrations of a toxic gas and translate this chemical signal into benign molecules identifiable by humans.

Living engineered materials were created by hosting the selected microorganisms in clay-based ceramics with open hierarchical porosity. Such clay-based porous ceramics were prepared from air bubble templates using a foaming technique previously reported in the literature. [38,39] In this technique, a wet foam is first created by incorporating air bubbles in a suspension of surface-hydrophobized clay particles. The adsorption of the hydrophized clay particles onto the surface of the air bubbles allows for direct drying and sintering of the wet foam into a ceramic scaffold with porosity of 80-90%.

Porous ceramics obtained via this foaming route display a hierarchical family of pores with sizes in the ranges 20-130 µm and 20-80 nm. [39] The macropores originate from the templating air bubbles, whereas the micropores arise from the interstices between the clay particles. Such porous structure has been shown to absorb water through capillary forces due to the favorable wetting of the clay surface. This leads to a wicking effect that can be exploited for thermal insulation or cooling of architectural elements. [39] Because the wet foams can also be 3D printed, this processing technique allows for the manufacturing of complex-shaped porous structures with three-level hierarchical porosity that are not accessible using conventional fabrication techniques (Figure 2b-d).

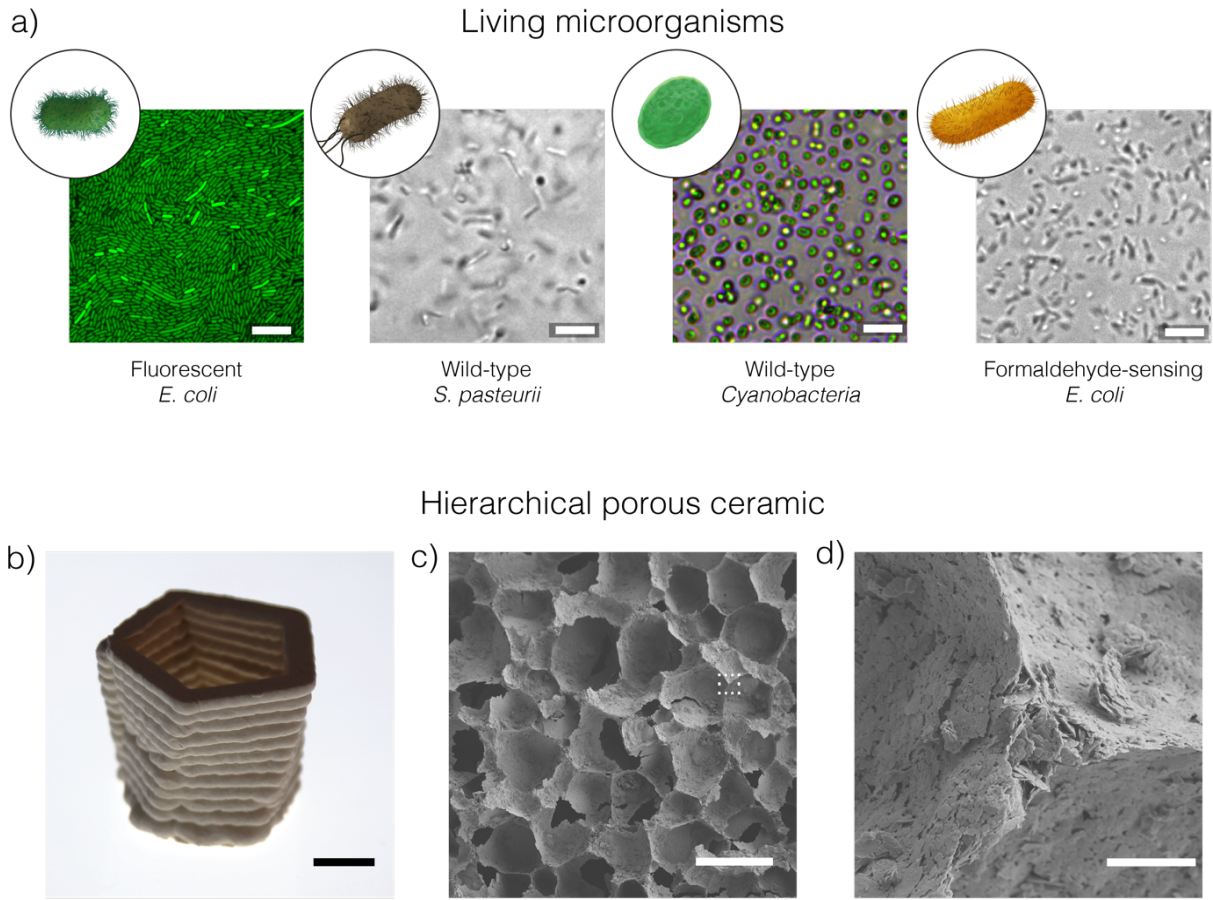

**Figure 2. Microorganisms and printed structures used for the creation of living porous ceramics.** (a) Illustrations and optical microscopy images of *E. coli* engineered to produce green fluorescent protein (GFP), the wild-type *Sporosarcina pasteurii*, the wild-type cyanobacteria *Synechococcus sp.*, and *E. coli* engineered to sense the hazardous gas formaldehyde. (b) 3D printed porous ceramic displaying (c) macropores and (d) micropores at different length scales. Scale bars: 10 µm in (a), 1 cm in (b), 100 $\mu$m in (c) and 5 $\mu$m in (d).

To provide a suitable home for microorganisms, porous ceramics need to fulfill three specific requirements. First, the porous structure should provide the nutrients and aqueous medium necessary for cellular growth and metabolic activity. Second, the selected bacteria must grow, colonize and remain metabolically active in the water-filled porous structure. Finally, the bacteria-laden porous ceramic needs to be sufficiently strong to withstand the mechanical loading expected in the intended application. The suitability of the foam-derived structures to fulfill these criteria was evaluated by conducting experiments that assess the hierarchical porous clay in terms of liquid uptake, bacterial growth and mechanical stability.

The ability of the porous structure to take-up liquid and to make this nutrient-rich medium available for bacterial growth depends on the wicking kinetics and the evaporation rate of the liquid inside the pores. The liquid medium will remain in the pores if the wicking rate throughout the structure is equal or higher than the evaporation rate across the surface of the monolith. To evaluate whether the porous ceramic satisfies this requirement, we measured the wicking and evaporation rates of aqueous medium through representative printed monoliths (Figure 3a,b). The measurements were performed on single and stacked monolithic units in contact with a water reservoir to mimic the possible arrangement of porous bricks in the envisioned living building material (Figure 1).

The experiments revealed that the transport of liquid through the tested monoliths is up to two orders of magnitude faster than the evaporation of the liquid medium across the sample surface (Figure 3a,b). For a single-unit monolith, wicking occurs predominantly within the first 5 minutes at an average rate of 47 mL h$^{-1}$ and follows the square-root time dependence expected from the Washburn model for wetting liquids. [39] By contrast, evaporation takes place at a constant rate in the range 0.32-0.76 mL h$^{-1}$ over 18 hours, which is typically observed when the process is controlled by a liquid film formed on the surface of the monolith. [40] These experimental data provide useful guidelines for the design of porous monoliths that can remain filled with nutrient-rich liquid, while allowing for constant liquid circulation driven by wicking (SI). Notably, wicking also happens across stacked monoliths, indicating that the liquid is able to effectively bridge between vertically adjacent units. The experiments show that the wicking rates for the initial water uptake decrease with the number of stacked monoliths, which suggests that mass transport across the liquid bridge becomes the rate-limiting step in such multi-unit systems.

Interestingly, after approximately 40% of the water capacity of the stacked monoliths system is reached, its weight continues to increase at a rate proportional to the number of additional monoliths (Figure S1). This phenomenon might be related to the lateral growth of the liquid bridge formed between monoliths, which enables continuous water uptake in 2- and 3-stacked monoliths beyond the initial imbibition phase. Such interpretation is supported by the fact that the single monolith (no liquid bridge) does not show further water uptake after the initial wicking. The lower rate of wicking observed for 2 and 3 stacks at this later stage might arise from the fact that the menisci of the liquid bridge formed between monoliths display a larger radius of curvature and therefore should grow at a lower rate compared to the menisci formed inside the porous monoliths.

The favorable wicking conditions enabled by the porous structure opens the possibility of infiltrating and colonizing the ceramic scaffold with microorganisms. To demonstrate this, we designed an experiment in which a culture medium loaded with GFP-producing bacteria is infiltrated into a stack of piled monoliths through the action of capillary forces (Figure 3c,d). The bacterial colonization of the scaffolds through this mechanism was quantified by measuring the total number of cells in the material over time using the most probable number method. [41] The results showed that the density of bacteria in the scaffold grows exponentially up to $10^6$-$10^7$ mL$^{-1}$ within the first 5 hours of the experiment. This suggests that the solid phase of the scaffold does not prevent the bacteria from proliferating as they would in a nutrient-rich liquid environment. Such proliferation behavior allows for extensive colonization of stacks of porous monoliths, as visually demonstrated by an experiment in which GFP-producing *E. coli* were used to infiltrate scaffolds through wicking and bacterial growth (Figure 3d).

While the high porosity of the scaffolds is crucial to enable bacterial colonization, it is also important that the porous structure is sufficiently strong to withstand the mechanical stresses expected during infiltration and final application. To evaluate this, we performed compression tests on porous cylindrical samples in three distinct representative conditions: fully dry, partially wet and fully wet (Figure 3e,f).

With an average compressive strength of 0.9 MPa and an average elastic modulus of 16 MPa (Figure S2), the fully dry scaffolds reached mechanical properties comparable to standard dense bricks. [42] The partially wet scaffold showed the lowest mechanical strength (0.7 MPa), which may result from the capillary forces that develop within the structure when water menisci are formed between the particles. This implies that the wicking process is a critical phenomenon for the mechanical integrity of the porous structure.

To ensure that the porous structure is not damaged by wicking-induced capillary forces, the ceramic scaffold can be further reinforced through a bacteria-driven biocementation process at ambient temperature. We demonstrate this reinforcing effect using wild-type *Sporosarcina pasteurii*, which are soil bacteria that can induce the precipitation of calcium carbonate in the structure if supplied with nutrients and calcium ions from a mineralizing aqueous solution (Figure 3g,h). By immersing a pre-sintered, 3D printed porous scaffold for 3 days in such a mineralizing solution, it is possible to increase the stiffness and the load-bearing capacity of the printed structure by, respectively, 55 and 78%, compared to an abiotic control sample (Figure 3g). Such strengthening effect correlates with the calcium carbonate content of the scaffold, which is, respectively, 9 and 42% in the abiotic and biotic sample after the biocementation process (Figure 3g, inset).

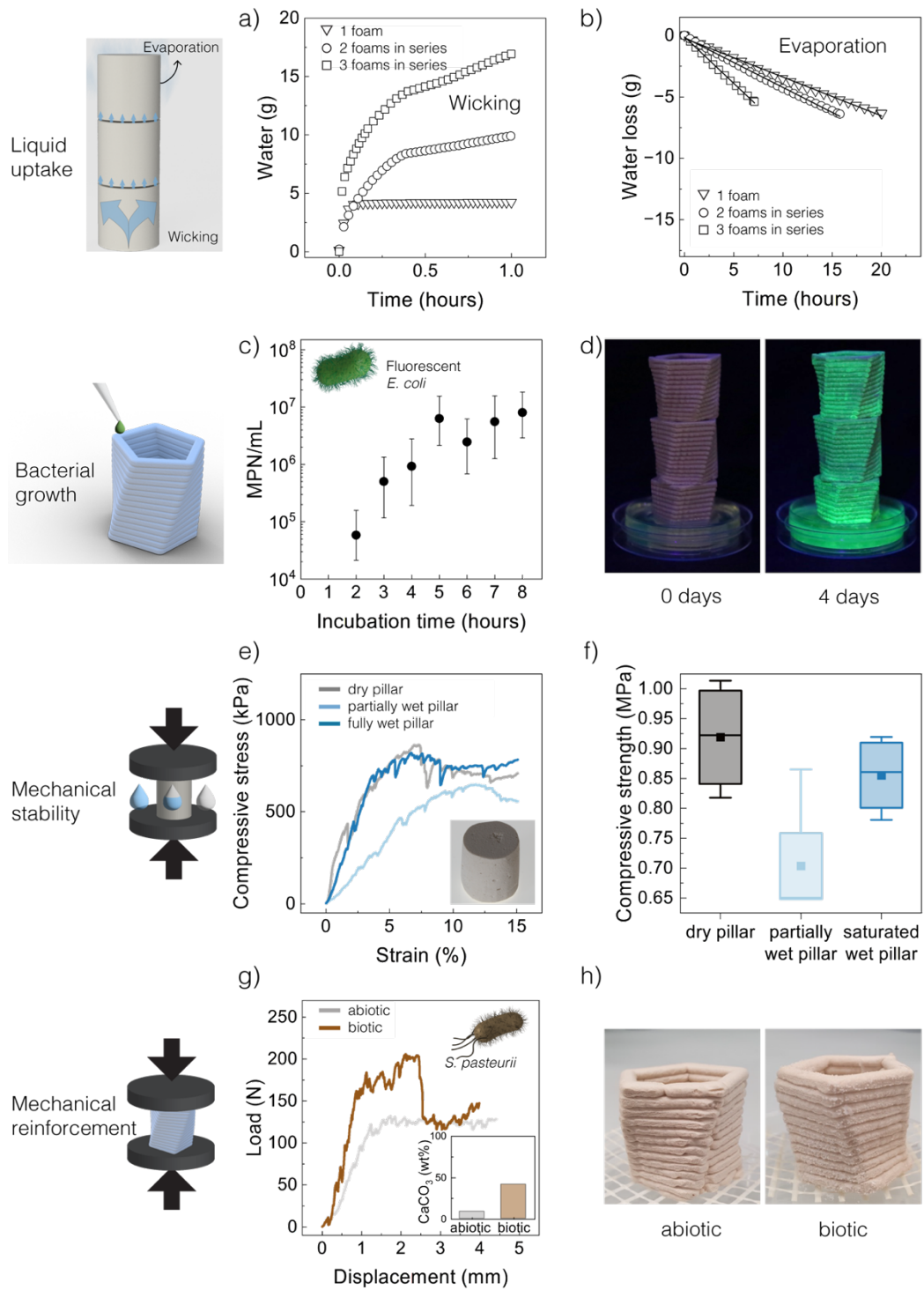

**Figure 3. Porous ceramics as scaffolds for metabolically active bacteria.** (a) Wicking and (b) evaporation behavior of the porous monoliths quantified by, respectively, the uptake and loss of water as a function of time. (c) Density of fluorescent *E. coli* in the porous scaffolds as a function of incubation time. (d) Photographs of stacked 3D printed monoliths under UV light, illustrating the homogeneous distribution of fluorescent bacteria (green) achieved after 4 days of incubation. (e,f) Compression tests performed on cylindrical scaffolds in the dry, partially wet and fully wet states. Plots (e) and (f) display representative stress-strain curves and box-plots of the compressive strength obtained for the tested samples. (g,h) Mechanical reinforcement of pre-sintered ceramic scaffolds by bacterial-induced biocementation. (g) Representative stress-strain response and carbonate fraction (inset) of biocemented scaffolds compared to a control sample without bacteria. (h) Photographs of the abiotic and biotic monoliths after biocementation for 3 days.

The possibility to infiltrate and grow microorganisms in ceramic scaffolds opens the way for the implementation of bacteria-regulated functionalities in the living material. To harness this potential, we studied the growth of cyanobacteria inside the ceramic scaffolds and evaluated the resulting living structure in terms of $CO_2$ capturing capabilities (Figure 4). To quantify the growth of the microorganisms into the porous ceramic, we measured the auto-fluorescence of the cyanobacteria in a customized experimental setup. In contrast to the fast-growing *E. coli* (Figure 3e), the proliferation of the cyanobacteria inside the scaffold was found to initiate only after 4 days of incubation (Figure 4a). This experimental result can be explained by the doubling time of these microorganisms, which is typically 20-40 minutes for *E. coli* and 2.6 hours for *Synechococcus sp.* [43,44]

The growth of cyanobacteria in the scaffolds is influenced by the porosity and pore size of the ceramic structure. To study this effect, we prepared porous ceramics with distinct porous architectures and measured the biomass generated in these scaffolds after bacterial growth. By combining various processing and sintering routes, three different porous architectures were prepared and investigated: (1) a low-porosity ceramic with 23% micropores, (2) a high-porosity scaffold with 90% macropores, and (3) a high-porosity hierarchical scaffold with both macro- and micropores (93% in total). After infiltration with cyanobacteria and incubation for 13 days under artificial light, the structures were characterized in terms of biomass generated through bacterial photosynthesis (Figure 4b).

The results revealed that the presence of macropores in the two high-porosity scaffolds increased the biomass content by a factor of 5 to 10 compared to the low-porosity structure containing only micropores. Moreover, the combination of micro- and macropores in the hierarchical scaffold led to a two-fold increase in biomass compared to the one-level macroporous structure, in spite of the relatively similar total porosity (90-93%). While further research is required to elucidate the mechanisms underlying this effect, we expect the hierarchical porosity of the scaffold to facilitate the supply of nutrients and bacterial growth. In such a hierarchical architecture, the micropores are expected to provide the wicking function needed for liquid transport, whereas the macropores serve as a reservoir of culture medium for bacterial colonization.

Despite the enhanced cell growth enabled by the hierarchical scaffold, the fraction of biomass synthesized by the microorganisms is significantly lower than the overall open porosity of the structure. This limited growth likely arises from the fact that cyanobacteria need light for photosynthesis, restricting biomass accumulation to the illuminated surfaces of the sample. To assess the bacterial spatial distribution across the scaffold, we cut a hierarchical sample after cell growth and inspected its cross-section in a fluorescence microscope. The auto-fluorescence of the chlorophyll under 625-650nm light confirmed the preferential growth of the cyanobacteria within a 1.2-2.0 mm layer close to the surface of the scaffold. Thicker layers of biomass were found close to large crevices of the sample, which should allow for deeper penetration of light into the scaffold and thereby improve bacterial photosynthesis.

To quantify the carbon capture effect resulting from the photosynthetic process, we measured the amount of $CO_2$ gas consumed by the bacteria-laden porous structure over time (Figure 4d-f). The $CO_2$ was quantified by taking aliquots of the gas phase in contact with the sample inside a customized closed chamber. In this analysis, a porous clay sample loaded with cyanobacteria was compared with a reference liquid sample containing *Synechococcus sp* at a cell density that leads to the same chlorophyll autofluorescence as in the monolith. The $CO_2$ concentration in gas aliquots taken at regular intervals was measured by gas chromatography over a period of 6 hours.

The $CO_2$ measurements showed that the photosynthetic activity of the microorganisms enabled effective carbon capture in the bacteria-laden scaffold. This is illustrated by the steady drop in the $CO_2$ concentration in the gas phase from 600-650 ppm to approximately 250 ppm within the first 5-6 hours of the experiment (Figure 4e). Carbon capture occurred 2.9-fold faster in the cell-laden scaffolds compared to the liquid culture of cyanobacteria (Figure 4f). Such an increase in carbon capture ability is partly related to the higher surface area of the scaffold relative to the liquid sample. Importantly, if the capturing rate is normalized over the envelope surface area, we still obtain a 1.7-fold faster carbon capture in the monolith compared to the liquid culture. This is attributed to the higher effective surface provided by the porous microstructure of the monolith. With a normalized carbon capture rate of 0.29 $\mu$mol m$^{-2}$ s$^{-1}$, the bacteria-laden ceramic scaffolds display a $CO_2$ consumption rate that is about 4 times higher than that recently reported for an microalgae-based living hydrogel. [32] By hosting the bacteria in a three-dimensional structure, the scaffolds provide means to increase the effective area of illuminated surface while keeping a footprint comparable to that of the liquid culture medium.

The ability to 3D print structures with complex three-dimensional shapes opens the possibility to further enhance the exposure of the cyanobacteria to light and thereby the carbon capture capability of the living porous ceramic. To explore this idea, we printed a channeled structure with a twisted geometry that combines high light exposure with the high accessible surface area provided by the hierarchical porosity (Figure 4g,h). When placed on top of bacteria-laden culture medium, the printed structure was effectively colonized by the cyanobacteria, which formed a smooth green biofilm on the exposed surface of the monolith (Figure 4i). Interestingly, we found that the bacteria proliferate particularly well on top of the grooves created between printed filaments, which provides another geometrical parameter to increase the carbon capture capabilities of the hierarchical porous structure.

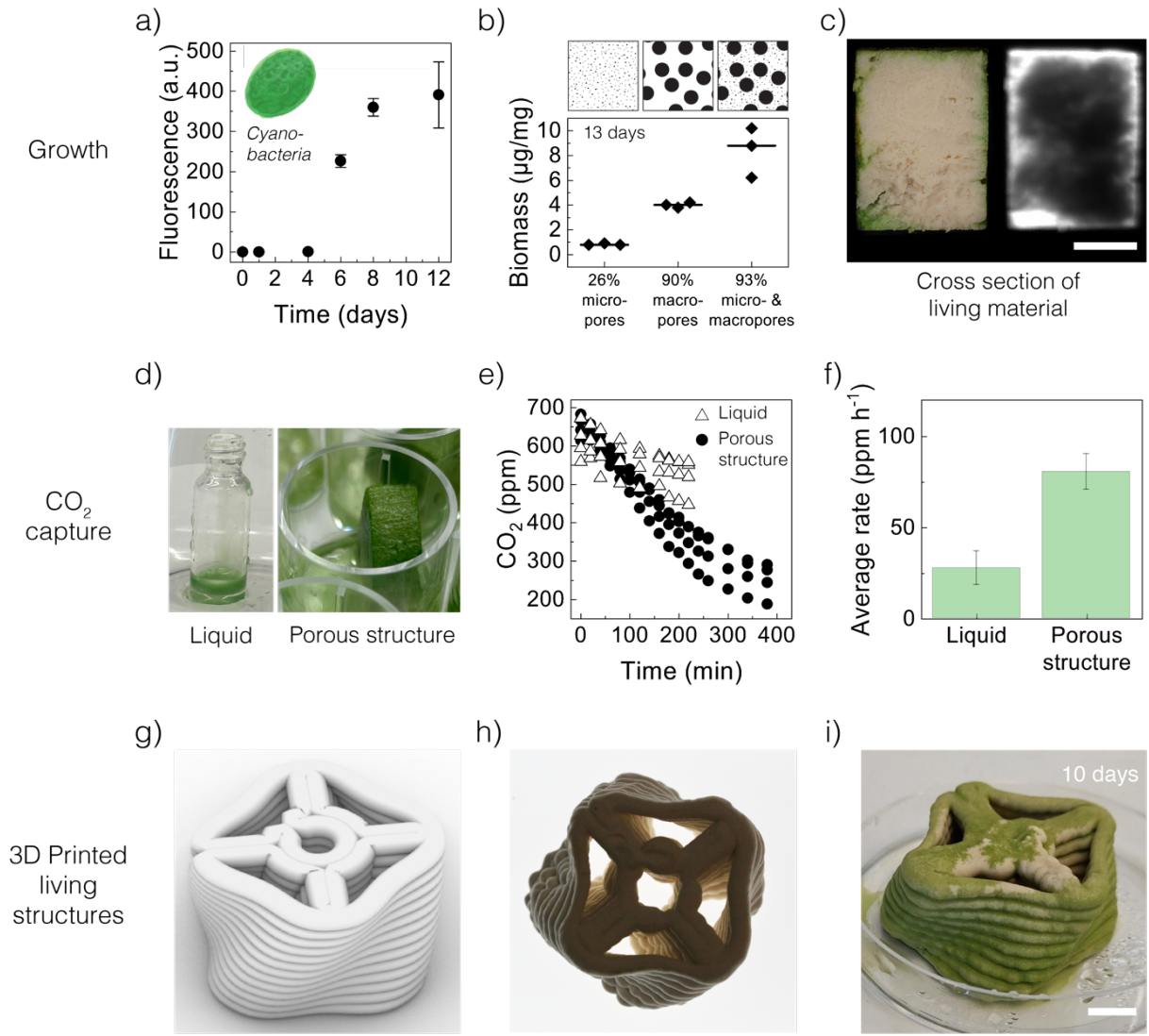

**Figure 4. Living porous ceramics with carbon-capturing cyanobacteria.** (a) Growth of cyanobacteria in the hierarchical porous ceramic quantified by the chlorophyll autofluorescence as a function of time. (b) Effect of the porous architecture on the biomass generated through bacterial photosynthesis. (c) Photograph (left) and fluorescence image (right) of the cross-section of a hierarchical porous ceramic after 8 days of bacteria growth. (d) Samples used to quantify the $CO_2$ capture ability of the bacteria-laden hierarchical porous monolith relative to a liquid culture. (e) Evolution of the $CO_2$ concentration in the gas phase during photosynthesis in porous and control samples. (f) Average rate of $CO_2$ capture measured for the hierarchical porous structure compared to the reference sample. Normalization with the envelope surface area gives 0.29 and 0.17 $\mu mol\ m^{-2}\ s^{-1}$ for the porous structure and the reference sample, respectively. (g-i) Rendering (g) and realization of a 3D printed hierarchical porous structure before (h) and after (i) cyanobacteria growth for 10 days. Scale bars: 5 mm in (c) and 10 mm in (i).

In addition to carbon capture, the hierarchical porous ceramics provide a suitable scaffold to host bacteria for other attractive functionalities, making use of the high porosity and mechanical stability of the structures. The spectrum of possible functionalities is vast, as the tools of synthetic biology can utilize and combine diverse biological functions through genetic engineering. To illustrate the potential of this approach, we designed and tested a bacteria-regulated living material that is able to detect toxic gases in the atmosphere and translate this information into a benign gas as an amplified signal that can be easily sensed by the human nose. Formaldehyde was chosen as an exemplary toxic gas that is a probable carcinogen for humans and is a well-known indoor pollutant. [45] Importantly, humans are only able to smell formaldehyde at concentrations around 1 ppm in air [46], which is 10 times higher than concentrations that are reported to be safe against sensory irritation, lung damage, and increase cancer risk. [47,48]

To allow humans to detect harmful formaldehyde concentrations below our olfactory capacity, we built a genetic circuit that triggers the engineered bacteria to convert isoamyl alcohol into isoamyl acetate in the presence of minor concentrations of the toxic gas (Figure 5a). Isoamyl acetate is a harmless room-temperature gas that has the scent of banana and is easily detected by humans at concentrations as low as 0.002 mM in alcohol-water mixtures and 0.002 ppm in the air. [49,50] This pathway has been suggested as an attractive olfactory reporter for biosensors, due to its general absence from the environment, low toxicity, and human sensitivity. [51]

The genetic construct designed for formaldehyde sensing was based on a metabolic pathway used by *E. coli* for formaldehyde detoxification. [52] In this pathway, the FrmR transcriptional repressor binds to its cognate promoter and thereby prevents transcription of formaldehyde-degrading enzymes *frmA* and *frmB*. When toxic formaldehyde is present, the formaldehyde molecule binds to the FrmR protein and induces a conformational change that detaches it from the promoter, thus allowing for transcription and detoxification. We based our designs on the variant promoter m4 from plasmid pTR47m4, which was found to have higher sensitivity to formaldehyde. [53] Previous work has shown that the genetic construct used for the detection of formaldehyde does not suffer from cross-sensitivity with methanol and two other aldehydes. [53] Our experiments confirmed the absence of cross-sensitivity with methanol and also with ethanol up to 1 mM (Figure S3). This high selectivity is a major advantage of the living sensor in comparison to commercially available sensors, which typically detect both ethanol and formaldehyde molecules indistinguishably. Preliminary experiments were conducted with a GFP-expressing construct pTR47m4-GFP, which optically confirmed the induction of the m4 promoter in engineered *E. coli* in the presence of formaldehyde gas (Figure S4).

In order to express a volatile output detectable by humans, we use the enzyme ATF1 from *Saccharomyces cerevisiae*. This enzyme produces isoamyl acetate (IAct), an odorous banana-smelling ester [54] through the esterification of isoamyl alcohol (IA) with acetyl coenzyme A. [55] Plasmid pFSKm4-ATF1 was built with the ATF1 sequence placed downstream of the m4 promoter, as well as constitutively expressed FrmR, so it would express ATF1 upon formaldehyde detection.

The formaldehyde-sensing capabilities of the engineered *E. coli* were probed by first using a liquid culture inoculated with bacteria at two different concentrations (Figure 5c,d). In this experiment, the culture medium initially contained 50 μM of formaldehyde and 10 mM of the precursor IA (Figure S5). Previous experiments had shown that these formaldehyde and IA concentrations enable the production of IAct only in the presence of formaldehyde, indicating efficient repression of ATF1 in the absence of the inducer (Figure S6). To assess the metabolic activity of the engineered bacteria, we measured the concentration of IAct produced and the concentration of IA consumed by the microorganisms over time.

The engineered activity of *E. coli* harboring pFSKm4-ATF1 in liquid medium led to a steady increase in isoamyl acetate and decrease in isoamyl alcohol over a period of 6 hours, indicating that the bacteria were able to effectively translate a weak input signal of formaldehyde (50 µM) into an amplified output signal of isoamyl acetate (up to 2.8 mM). By increasing the initial bacterial concentration by 50-fold, from $1 \times 10^7$ to $5 \times 10^8$ CFU mL$^{-1}$, we observed a 22-fold increase in the average conversion rate within the first 3 hours of the experiment (Figure S7). This indicates the importance of creating scaffolds capable of hosting high bacterial concentrations to enable the detection and amplification of the formaldehyde molecule through the genetic construct.

To implement this bacteria-assisted sensing capability in three-dimensional structures, we colonized a hierarchical porous scaffold with the engineered *E. coli* and evaluated its ability to generate isoamyl acetate in the presence of formaldehyde in the gaseous state. In contrast to the liquid culture tests, these experiments were conducted in a customized setup to capture the conditions expected in a real application, where formaldehyde and isoamyl acetate should be detected as gases in the environment around the living scaffold (Figure S8). In this experiment, the performance of the bacteria-laden porous scaffold was compared with that of a dense brick and a liquid culture containing the same concentration of bacteria.

Porous scaffolds colonized with the engineered *E. coli* were able to sense 0.12 ppm of formaldehyde gas and convert this low input signal into a human-detectable level of isoamyl acetate (41 ppm) after three hours of exposure to the toxic gas (Figure S8). Importantly, this formaldehyde concentration is 8-times lower than the odor threshold limit that can be detected by humans [46] and comparable to the safe dose in terms of irritation and cancer hazards (0.1ppm). [47] By contrast, the dense brick and the liquid culture generated significantly less isoamyl acetate when exposed to the same levels of formaldehyde (Figure 5e), which is most likely due to their lower air-liquid interfacial area compared to the porous scaffold. The ability to host a high concentration of bacteria and distribute them over a large accessible area makes the living porous ceramic a promising platform for gas sensing of hazardous gases at room temperature with minimal energy input.

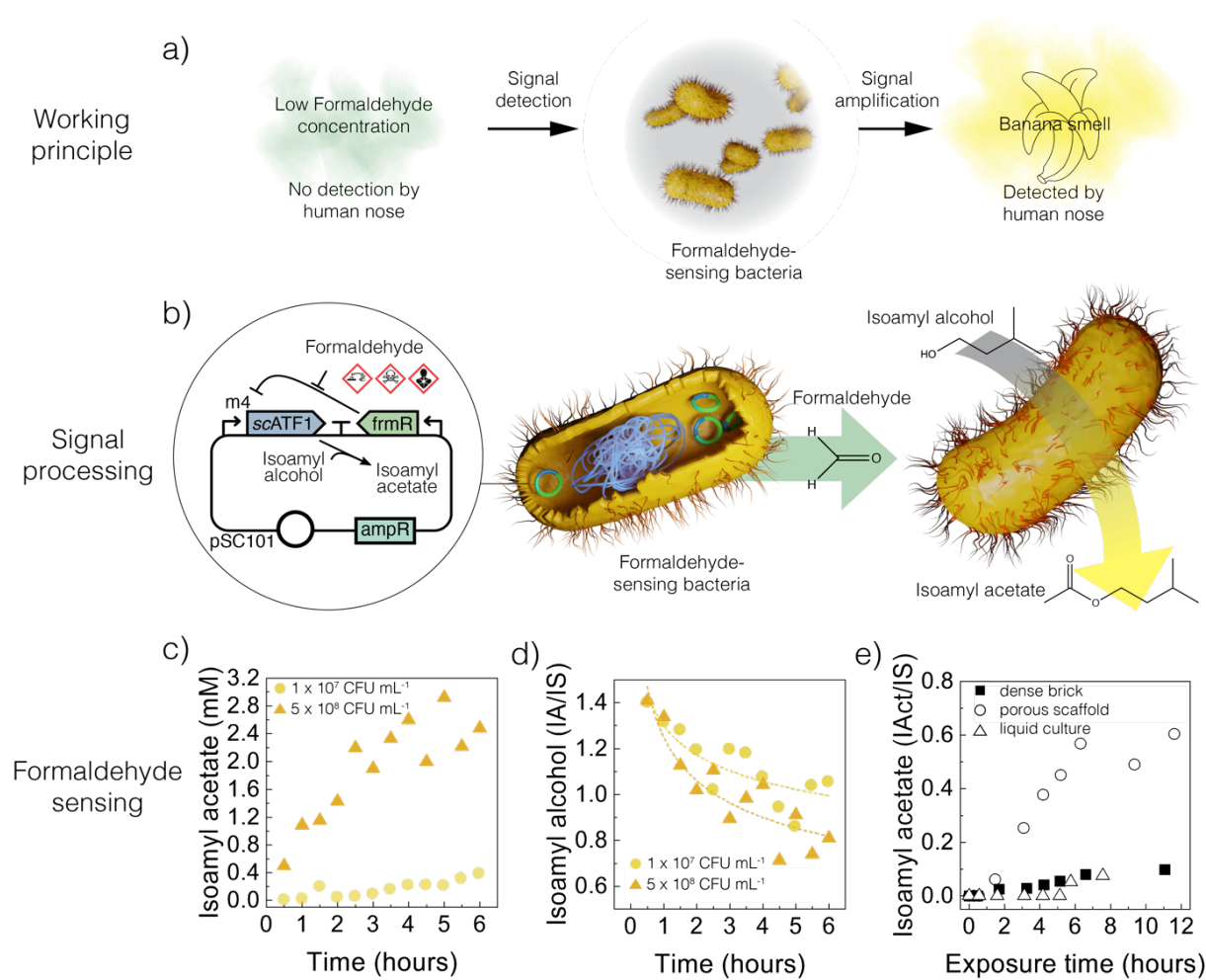

**Figure 5. Living porous ceramic for sensing of gaseous formaldehyde.** (a) Working principle of the bacteria-assisted sensing mechanism, highlighting the detection of a low concentration of toxic formaldehyde and its conversion into an amplified signal of an human-detectable scent. (b) Plasmid incorporated in the engineered bacteria to express isoamyl acetate from isoamyl alcohol triggered by the presence of formaldehyde. (c,d) Concentrations of (c) isoamyl acetate generated and (d) isoamyl alcohol consumed over time in culture media containing 50 µM formaldehyde and an initial concentration of formaldehyde-sensing *E. coli* of $1 \times 10^7$ or $5 \times 10^8$ CFU mL$^{-1}$. (e) Evolution of isoamyl acetate produced by bacteria inoculated in a hierarchical porous scaffold, in a dense brick or in liquid culture exposed to 1.2 ppm formaldehyde in air. The concentrations displayed in plots (d) and (e) were normalized with respect to the concentration of the internal standard (IS) 1-propanol.

**Conclusions**

The colonization of porous ceramics by metabolically active microorganisms enables the creation of engineered living materials with enticing carbon-capture and gas-sensing functionalities. Carbon capture is achieved by hosting wild-type photosynthetic cyanobacteria within the pores of the ceramic scaffold. This enabled the fabrication of living porous structures that can remove $CO_2$ directly from air and store it as organic matter inside the cyanobacteria. For gas-sensing, the colonizing bacteria are genetically engineered to detect toxic formaldehyde gas and transduce this chemical signal into an amplified banana scent detectably by humans. To process chemical information in a highly selective manner, the bacteria are modified with a genetic construct that is specifically designed to detect, translate and amplify the input signal into a specific output gas. By providing room for microbial growth and transporting nutrient-rich water, the hierarchical porosity of the ceramic scaffold is crucial for the proliferation and metabolism of such functional microorganisms. The transport of nutrient-rich water to the microorganisms is driven by capillary forces through a spontaneous wicking mechanism that does not require external energy input for operation. The ability to transport water autonomously and to process chemical information in a programmable, self-regulated manner makes bacteria-laden porous ceramics a powerful platform for the design and creation of functional engineered living materials.

**Acknowledgements**

We thank ETH Zurich for the financial support. The authors would also like to acknowledge Dalia Dranseike and Barna Benke for sharing the chlorophyll assay, David Kiwic for the introduction to the $CO_2$ quantification setup, Marco Binelli for the renders of the bacteria, and Karen Andrea Antorveza Paez for helping in designing the 3D printed structures. We are also grateful to Mathias Steinacher and Elena Tervoort for the fruitful discussions.

**Author contributions**

The concept of this study was developed by the Alessandro Dutto, Anton Kan and André R. Studart. The porous ceramic monoliths and prints were prepared by the Alessandro Dutto, Zoubeir Saraw and Aline Maillard. Zoubeir Saraw conducted the water uptake, bacterial proliferation and mechanical studies. Alessandro Dutto and Aline Maillard investigated the growth and carbon capturing capability of the cyanobacteria. Alessandro Dutto and Daniel Zindel developed the methods for the gas sensing setup. Anton Kan engineered the gas sensing E. coli and performed the tests for toxicity and cross-sensitivity. The figures were prepared by Alessandro Dutto and revised by André R. Studart. The text was written by André R. Studart and Alessandro Dutto and was revised by all authors.

**Materials and methods**

**Preparation of porous ceramics**

Porous ceramics were prepared from wet foams stabilized by clay particles.[39] The particle-stabilized foams were generated from 35 wt% suspensions of clay (WM, Sibelco) in a water-glycerol mixture, which was supplemented with foaming additives provided by FenX AG. The suspension was homogenized using a planetary mixer (ARE-250, Thinky) and mechanically frothed in a 100 mL beaker for 3 minutes using a kitchen mixer (Multimix 5, 750W, Braun). The resulting viscoelastic wet foam was either casted into pillars or filled into a 50 mL syringe (BD Plastipak, BD) for 3D printing by direct ink writing (DIW).

Twisted hexagon and twisted trunk structures were 3D printed using a custom modified desktop printer (Ultimaker 2+). The original print head was replaced by a custom-made extrusion system consisting of a mechanically driven syringe pump that can accommodate 50 mL syringes.[8] The print paths were designed using the software Grasshopper (Rhinoceros, Robert McNeel & Associates). A nozzle with a diameter of 1.6 mm was used to print objects with typical individual layer height of 3 mm at a print-head velocity of 40 mm s$^{-1}$, and an extrusion flow rate of 30 mL min$^{-1}$.

The printed and casted parts were initially dried at room temperature for a period of one day, after which they were transferred to a 60°C ventilated oven where they were dried overnight. The dried samples were then transferred to a setter plate and fired in an electrical furnace (HT 08/17, Nabertherm) using a multi-step heating schedule. The heat treatment comprised a burnout period (200°C for 2 hours, 290°C for 2 hours, and 400°C for 1 hour) and a sintering period (1000°C for 2 hours). The heating rate was fixed at 1°C/min for the burnout period and 4°C/min for the sintering period. A second sintering step was employed to obtain denser porous ceramics. For this, the burnout period was skipped, and the ceramic was heated to a peak sintering temperature of 1150°C at a heating rate of 5°C/min and held for one hour. The cooling rate was given by the heat loss of the oven and was not actively controlled.

**Water uptake and evaporation**

To quantify the liquid flow within and across the porous ceramics, wicking experiments were performed on sintered pillars of approximately 2.6 cm and 2 cm in height and diameter, respectively. The pillars were ground on their top and bottom surfaces to ensure sufficient open porosity for wicking and evaporation of liquids. The experiments were conducted using distilled (Milli-Q) water. During the test, the weight of the pillars was measured using the data collection software RsMulti (A&D Company, Limited, Japan) with an analytical balance (XS204, Mettler Toledo) equipped with a density kit (Mettler Toledo). For the wicking experiments, 1, 2 or 3 pillars were stacked on each other and the bottom of the stack was submerged in distilled (Milli-Q) water up to a depth of around 2 mm. This immersion depth was kept constant by supplying the water reservoir with additional water at a rate of 2 mL h$^{-1}$. For the evaporation experiments, the weight of the pillars was tracked following the same procedure, starting from fully wet samples.

**Mechanical properties**

The mechanical stability of the porous ceramics at different moisture levels was evaluated by performing uniaxial compression tests on sintered pillars with a diameter and height of 2.6 and 2 cm, respectively. The compression tests were conducted on a universal testing machine (AGS-X, 1 kN load cell, Shimadzu) at a cross-head speed corresponding to a strain rate of 0.10 per minute. Prior to testing, the samples were ground on the top and base with sandpaper (SiC, CIMI 1000) to achieve a flat surface. The samples were pre-conditioned before the test to be dry, partially wet, or completely wet. Dry pillars were subjected to a ventilated oven at 60°C for a minimum of three hours. Partially wet pillars were obtained by allowing the dry pillars to wick up deionized water for 30 minutes, reaching 70–80% of their total liquid imbibition capacity. Completely wet pillars were prepared by applying vacuum to the wicked samples, so that the porosity could be completely filled with the liquid.

**Bacterial growth**

Distinct culture media were prepared for the growth of microorganisms. An urea yeast extract medium was used to grow *Sporosarcina pasteurii*, whereas LB and HSC media were utilized to cultivate *Escherichia coli* and *Synechococcus sp.* PCC 7002, respectively.

The urea yeast extract (UY) medium was prepared by adding 30 g L$^{-1}$ urea (≥ 99.0%, ACS reagent) and 20 g L$^{-1}$ yeast extract (YE, Millipore) in 0.13 M TRIS buffer (tris(hydroxymethyl)aminomethane, 1.0 M buffer solution at pH 9.0, Thermo Fischer). The buffer was previously diluted with deionized (Milli-Q) water and had its pH adjusted to 7 using concentrated hydrochloric acid (37%, VWR). The media was prepared by mixing a 2x concentrated urea solution and a 2x concentrated yeast extract solution in a 1:1 ratio. The urea and yeast extract solutions had been previously sterilized by filtering and autoclaving, respectively. LB media was prepared from 20 g L$^{-1}$ LB-broth Lennox (Sigma-Aldrich) and 5 g L$^{-1}$ sodium chloride (VWR chemicals) using deionized (Milli-Q) water. High salt concentrated (HSC) media was prepared as described in the SI.

Working cultures of *Sporosarcina pasteurii* (DSM 33, Leibniz Institute DSMZ) were inoculated from a glycerol stock and grown in UY medium at 30°C under shaking conditions (200 rpm) in an incubator (Minitron, Infors HT). Green fluorescent protein (GFP) *Escherichia coli* (DH5α bearing plasmid pKAG) was inoculated from a single colony and grown in LB media supplemented with 50 μg mL$^{-1}$ kanamycin at 37°C under shaking conditions (200 rpm). Formaldehyde sensing *Escherichia coli* (DH5α bearing plasmid pFSKm4-ATF1) was inoculated from a single colony and grown in 1 mL LB media supplemented with 100 μg mL$^{-1}$ carbenicillin at 37°C under shaking conditions (200 rpm). 100 μL of this pre-culture were typically added to 10 mL media and grown overnight. *Synechococcus* sp. PCC 7002 (referred to hereafter as cyanobacteria) was cultivated in HSC media at 30°C under 16/8 h day/night cycles and shaking conditions (150 rpm). The light intensity during the day was set to 48 μmol m$^{-2}$ s$^{-1}$. The cyanobacteria cultures were refreshed every 10 days by replacing 2/3 of the culture with fresh HSC media.

To assess the stage of bacterial growth in the media, the optical absorbance of the liquid cultures was measured in 96-well plates using a microplate reader (Varioskan Lux, Thermo Scientific). The absorbance was acquired at 600 nm for *S. pasteurii* and *E. coli*, and at 700 nm for the cyanobacteria. OD$_{600}$ was calculated by pathlength correction in the Varioskan software.

**Inoculation of ceramics with microorganisms**

Different inoculation techniques were tested to achieve high reproducibility in terms of homogeneous cell growth. The best techniques were found to be simple dipping in an overnight culture for *E. coli* or *S. pasteurii*, and dipping in a running culture in the case of the cyanobacteria. Prior to inoculation, the porous ceramic samples were first imbibed in fresh sterile media under vacuum to ensure that all the accessible porosity was filled with the liquid. Subsequently, the porous ceramics were inoculated by dipping them in the corresponding running culture. For incubation, the infiltrated samples were placed on top of glass beads with a diameter of approximately 5 mm, which acted as spacers to allow exposure of the bottom of the sample to the fresh media.

**Bacterial growth in porous ceramics**

The colonization of the porous ceramics by the microorganisms was assessed following different protocols depending on the bacteria used.

The growth and viability of *E. coli* and *S. pasteurii* within the porous ceramics was quantified by estimating the concentration of viable bacteria in the samples using the most probable number (MPN) method. A pre-treatment protocol similar to that reported by Lee *et al* [41] was developed to collect bacteria from the porous ceramic structures. For this, small portions of the structures were crushed with a spatula in a tube containing a sterile solution of 0.85 wt% sodium chloride. The amount of solution was 10 times the weight of the tested sample. After vortexing at 2'700 rpm for 5 minutes (Vortex-Genie 2, Scientific Industries Inc.), the tubes were sonicated in an ultrasonic bath for 6 minutes (130 W, Branson 1510, Branson Ultrasonics Corporation). Ice was added to the bath to prevent overheating. After sonication, the suspension was centrifuged at 1400 $g$ for 15 minutes. The supernatant obtained from the centrifugation step was typically diluted to $10^{-2}$ to obtain sufficiently low bacteria counts and thereby reliable results with the MPN method. 20 μL aliquots of the resulting bacteria solutions were added to a 96-well plate containing 180 μL of the adequate growth medium. Abiotic negative controls were prepared by replacing the bacteria solution with 20 μL of sterile 0.85 wt% sodium chloride solution. After serial dilutions of 1:10 up to the 8th row, the plates were incubated overnight under different conditions depending on the microorganism. After overnight incubation, positive and negative wells were scored. A well was considered positive if it became turbid after incubation. Typically, positive wells had an optical density ($OD_{600}$) greater than 0.1. Colony forming units were reported as MPN mL$^{-1}$ with 95% confidence intervals. Statistical calculations of the MPN and corresponding confidence intervals were performed using an on-line tool developed by the United States Environmental Protection Agency (EPA) [56]. The input to the calculator includes the number of dilutions, the number of replicates for each sample, the amount of sample used in the most concentrated dilution, and the combination of positive and negative wells for each dilution. For the best accuracy and comparability, the number of dilutions included in the calculations must to be consistent. In this work, three dilutions and a minimum of three replicates per sample were always used for the calculations.

The growth of the cyanobacteria within the porous structures was evaluated using a chlorophyll assay. [57] To extract the chlorophyll, the living porous ceramic was transferred to a Falcon tube containing 1, 2 or 3 mL methanol (≥99.9%, Sigma Aldrich). The resulting suspensions were shortly vortexed at 2'700 rpm and afterwards incubated at room temperature for 1 hour under shaking conditions at 140 rpm (Rotamax 120, Heidolph). After incubation, the porous ceramics had a pale appearance whereas the supernatant was green. Notably, structures older than 4 days required more methanol (2 or 3 mL) to extract all chlorophyll. The addition of more methanol was later considered as a dilution factor in the calculations. To quantify the amount of chlorophyll extracted from the porous ceramics, the supernatant was passed through a 0.2 μm syringe filter and the autofluorescence of the chlorophyll was measured using a plate reader in a pigmented 96-well plate at excitation and emission wavelengths of 435 and 675 nm, respectively. The chlorophyll content of cyanobacteria cultured in media were also measured to serve as a reference. For this purpose, 100 μL of the culture was added to 1 mL of methanol and the above procedure was repeated.

**Biocementation**

Porous ceramics were mechanically reinforced through the bacteria-induced calcification of the structure in water at room temperature. Calcification was achieved through the precipitation of calcium carbonate within the porous structure. In order to quantify the reinforcing effect of this biocementation process on the printed porous ceramics, abiotic and biotic twisted pentagon prints were subjected to a calcification step.

For calcification, the biotic structures were first inoculated and cultivated with *S. pasteurii* until an $OD_{600}$ of at least 2 was reached. Following this initial cultivation step, the structures were submerged in a 40 mL calcification bath composed of 1 M calcium chloride (≥93.0%, Sigma Aldrich) and 0.5 M urea (≥ 99.0%, ACS reagent) in 0.13 M TRIS buffer at pH 7. After a calcification period of 72 hours at 28 °C, the structures were removed from the bath and left to dry overnight in a ventilated oven at 60°C. Abiotic structures were prepared following the same protocol but without the initial inoculation step. Compression tests were carried out on both set of samples in accordance with the methodology described above.

To interpret the mechanical properties of the biocemented ceramics, the amount of calcium carbonate precipitated at the end of the calcification process was also measured. The mass of precipitated calcium carbonate was quantified by performing thermogravimetric analysis (TGA 5500, TA Instruments) of a representative portion of an abiotic and a biotic structure. The samples were grinded in a mortar to a fine powder and heat treated in air according to the following schedule: heating from 40°C to 500°C at a rate of 10°C min$^{-1}$, holding at 500°C for 1 hour, heating to 750°C at a rate of 5°C/min, and holding at 750°C for 10 minutes before cooling down. The mass of $CaCO_3$ in the samples ($w_{CaCO_3}$) was determined from the mass loss between 570°C and 750°C ($\Delta w_{570-750°C}$), which is the temperature range over which $CaCO_3$ is expected to thermally decompose into CaO and $CO_2$. Taking into account the stoichiometry of this reaction, the following equation was used: $w_{CaCO_3} = \Delta w_{570-750°C} \cdot \frac{M_w(CaCO_3)}{M_w(CO_2)}$, where $M_w(CaCO_3)$ and $M_w(CO_2)$ are the molar mass of $CaCO_3$ and $CO_2$, respectively.

**Sample imaging**

Macroscopic samples were photographed using a reflex camera equipped with a macro 100 mm objective (Canon EOS 6D) at varying illumination conditions. To image the GFP *E. coli*, a UV torch was employed to illuminate the samples.

Cross-sectional images of the porous ceramic colonized by the cyanobacteria were taken using a digital microscope (VHX-6000, Keyence). To capture the autofluorescence of the cyanobacteria, the cross-sections were also imaged with the ChemiDoc MP imager (Bio-Rad) under red epi illumination and a 700/50 emission filter.

Electron microscopy micrographs of the porous ceramics were recorded using a field emission electron microscope (GeminiSEM 450, Zeiss) in secondary electron mode at an acceleration voltage of 3 kV, a working distance of 7.5 mm, and a probing current of 100 pA.

**$CO_2$ capture**

The $CO_2$ capturing capability of the cyanobacteria was quantified by measuring $CO_2$ concentrations from a sealed flask over time using a gas chromatographer. For this purpose, cyanobacteria cultured for 12 days on the porous ceramic and in a liquid culture were first placed inside Erlenmeyer flasks and sealed with a rubber septum. To ensure a comparable cell density in the liquid culture and in the porous ceramic, the chlorophyll content of the living porous structure was determined following the protocol described above. Based on this information, the chlorophyll content and the volume of the liquid culture was made comparable to that of the porous ceramic. For both conditions, four replications were measured every 20 minutes. For this, the sampling volume of 4 mL was extracted from the overhead space of the flask using a 5 mL syringe. A volume of 0.5 mL was discarded, and the remaining 3.5 mL were manually injected into a gas chromatograph (Micro-GC 3000A) equipped with a thermal conductivity detector, using He and Ar as carrier gases. To prevent the formation of an underpressure in the flask, the sampled volume was re-injected with nitrogen gas. The dilution effect arising from the addition of nitrogen was taken into account to determine the actual $CO_2$ amount captured by the bacteria.

The impact of porosity on the growth of carbon-capturing cyanobacteria was quantified by comparing small ceramic samples with different porosities prepared from the same clay. These included a relatively dense cube subjected to a low sintering temperature (1000°C), a porous cylinder treated at a high sintering temperature (1150°C) and a porous cylinder subjected to a low sintering temperature (1000°C). Such heat treatments led to porosities of 26%, 90% and 93%, for the dense cube, the porous cylinder sintered at 1150°C and the porous cylinder sintered at 1000°C, respectively. To enable bacteria colonization, the samples were inoculated and incubated for 13 days. The biomass of the resulting living structures was quantified from the loss of ignition (LOI) of the dried samples. The LOI corresponds to the amount of organic mass lost by the sample during thermal treatment. To measure the LOI, the samples were heated to 550°C at a rate of 4°C min$^{-1}$ and held at this peak temperature for 5 hours. A high-precision scale (0.01 mg readability, XP205, Mettler Toledo) was employed to determine the weight loss of the heat treated samples.

**Plasmid Design and Construction**

Plasmid pFSKm4-ATF1 was based on the backbone of pTR47m4-GFP (Addgene #102436) [53], containing the pSC101 origin of replication, ampicillin resistance and the repressor *frmR* under the regulation of the *ptet* promoter, which was constitutive since the host *E. coli* DH5a lacked a *tetR* repressor. The ATF1 sequence was obtained from plasmid p006-Banana-late (Addgene #112251), [58] and included the moderate ribosome binding site BBa_R0064. This ATF1 sequence was used to replace the RBS and coding sequence of sfGFP on plasmid pTR47m4-GFP to create plasmid pFSKm4-ATF1. Physical DNA and full sequence information of pFSKm4-ATF1 can be found at Addgene.

DNA construction was performed using Gibson assembly, (Gibson 2008) following the protocol in reference using enzymes Taq ligase (M0208S), T5 Exonuclease (M0663S), and Phusion polymerase (M0530S), obtained from New England Biolabs (NEB). PCR fragments were obtained using custom DNA oligos (Integrated DNA Technologies) on the templates specified, designed to contain 25-30 bp overlaps between assembled fragments. PCR was performed using the Q5 ® Hot Start High-Fidelity (M0494S) from NEB. PCR fragments were purified with Reliaprep DNA Clean-Up and concentration system (A2892, Promega). Cloning was performed in E. coli DH5a chemically competent cells. Plasmid purification was performed with the PureYield Plasmid Miniprep System (A1223, Promega). Plasmid DNA sequence was confirmed by Sanger sequencing carried out by MicroSynth, Switzerland.

**Formaldehyde sensing**

The ability of the engineered *E. coli* to sense formaldehyde was evaluated by means of headspace gas chromatography-mass spectrometry (GC-MS) (Trace 1300 and ISQ, Thermo Scientific). Because the formaldehyde sensing capability is expected to depend on bacterial density, we first prepared experiments with liquid cultures inoculated with controlled initial concentrations of engineered *E. coli*. Inoculum cultures were prepared by picking and growing colonies of *E. coli* bearing pFSKm4-ATF1 overnight in LB media with 100 μg mL$^{-1}$ carbenicillin at 37°C under shaking conditions. This yielded turbid cultures with an OD$_{600}$ of 2.1, which corresponded to 1.1 x 10$^9$ CFU mL$^{-1}$. Cells from overnight cultures were harvested by centrifugation and redispersed at varying proportions in fresh LB media supplemented with 100 μg mL$^{-1}$ carbenicillin and 10 mM isoamyl alcohol (>99.0%, Fluka). This concentration of isoamyl alcohol was found to be a good compromise between not inhibiting bacterial growth and providing enough substrate for ATF1 (Figure S9). In the presence of formaldehyde, the engineered *E. coli* converts isoamyl alcohol into isoamyl acetate.

The impact of bacterial density on the resulting isoamyl acetate concentration was evaluated by the addition of 50 μM formaldehyde to freshly redispersed cultures inoculated with either 1 x 10$^7$ or 5 x 10$^8$ CFU mL$^{-1}$ respectively. During incubation of the cultures at 30°C and 200 rpm, small aliquots (200 μL) were sampled at given time intervals and placed in 10 mL headspace vials. 10 μL 1-propanol (99+%, Acros organics) was added to the vials before sealing to quench the bacteria activity and to serve as an internal standard for the quantification with GC-MS. The vials were thoroughly mixed, sonicated for one minute and left to rest for 15 minutes before 0.5 mL of the headspace was sampled and injected into the gas chromatographer. To quantify the isoamyl acetate concentration, the chromatogram was evaluated at selected masses at which isoamyl acetate, isoamyl alcohol and 1-propanol peaks are expected (*m/z* = 30 and *m/z* = 70). The ratio of the isoamyl alcohol and isoamyl acetate peak integrals to the 1-propanol peak integral (internal standard) was calculated and used for quantification. The isoamyl acetate concentrations were obtained from a calibration curve (Figure S5).

To evaluate the formaldehyde-sensing ability of the engineered *E. coli* inside living ceramics, we designed experiments with dense or porous samples using a liquid culture as a control. The dense and porous samples featured porosities of, respectively, 26 and 93%, and were prepared in the same as those used for the growth of cyanobacteria. To colonize the porous and dense structures with *E. coli*, samples were dipped into LB media (with 10 µg mL$^{-1}$ carbenicillin and 10 mM isoamyl alcohol) inoculated with 1 x 10$^9$ CFU mL$^{-1}$ and infiltrated by applying a vacuum for 20 seconds. The volume of the liquid culture used as control was the same as the one wicked by the porous ceramic. For the quantification of the sensing ability of the living ceramics, the samples were exposed to formaldehyde through the gas phase and gas aliquots were extracted for GC-MS measurements. 3 mL of aqueous solutions with 1.2 and 12 mM formaldehyde and 50 or 150 µL 1-propanol as internal standard was added to an Erlenmeyer flask into which a smaller vial containing the sample was placed. Using Henry's law, the equilibrium concentrations of formaldehyde in the gas phase around these solutions was estimated to be 125 and 1250 ppb. All samples were sealed with a rubber septum and left still at room temperature during collection of gas aliquots. At given time intervals, an aliquot of 0.5 mL was then taken from the headspace of the Erlenmeyer flask and injected into the GC-MS. In the absence of a calibration curve for all samples, the isoamyl acetate concentrations are reported relative to the internal standard (Figure S8).

# Supplementary information

## Living porous ceramics for bacteria-regulated gas sensing and carbon capture

Alessandro Dutto, [1] Anton Kan, [1] Zoubeir Saraw, [1] Aline Maillard, [1] Daniel Zindel, [2] André R. Studart [1]

[1] Complex Materials, Department of Materials, ETH Zürich, 8093 Zürich, Switzerland

[2] Laboratory of Physical Chemistry, ETH Zürich, 8093 Zürich, Switzerland

## Supplementary text

### Estimation of water transport in porous structures

In analogy to the water transport in trees, the transport of liquid through the pores of a partially wetted porous object is driven by the evaporation at the surface of the object and the cohesion forces of water. Molecular cohesion enables the transport of water over long distances within porous structures through the action of capillary forces. Such forces allow for the vertical transport of water even against the action of gravity. Surface evaporation from a liquid-saturated porous object is often the rate-limiting step that controls the speed of liquid transport through the structure. Here, we estimate the maximum vertical distance that water would be able to travel in our porous ceramics and the expected travelled distance in these structures in a possible application.

To estimate the maximum vertical distance travelled by water inside the porous ceramics, we use the idealized model of a vertical tube partially filled with a wetting liquid. Assuming a tube of radius $r_0$, the vertical distance ($h$) travelled by water in the tube through capillary action can be estimated by applying Jurin's law: [1,2]

$$h = \frac{2\gamma \cos\theta}{\rho g r_0} \tag{1}$$

where $\gamma$ is the surface tension of the wetting liquid, $\theta$ is the contact angle of the liquid on the solid wall of the tube, $\rho$ is the density of the wetting liquid and $g$ is the gravitational acceleration constant ($g = 9.81 \frac{m}{s^2}$). Taking water as the wetting liquid and considering the hydrophilic nature of the clay, we assume $\gamma = 72 \frac{mN}{m}$, $\theta = 10°$, and $\rho = 1000 \frac{kg}{m^3}$. The radius, $r_0$, can be approximated by the size of the micropores of the porous ceramic, which typically ranges from $20$ to $80\ nm$. [3] Provided that these pores are interconnected, Jurin's equation results in theoretical vertical distances ranging from $180$ m to $720$ m. This estimation clearly shows that the transport of liquid in such porous structures is dominated by surface forces rather than gravity.

While capillary forces enable water transport over very large vertical distances, it is important to note that the timescale for transport is also strongly dependent on the characteristic size of the pores. Such dependence of the wicking timescale on pore size is described by Washburn's equation:

$$t = h^2 \frac{2\eta}{r_0 \gamma \cos\theta} \tag{2}$$

where $\eta$ is the viscosity of the liquid. For water, $\eta = 0.89\ mPa \cdot s$. From this equation, one can predict that for pore sizes of 20-80 nm it would take 1 year for liquid to be transported along 5-10 m. This estimation indicates that the infiltration of the porous structure through immersion in liquid is the most time-effective approach to saturate the pores with the liquid phase. After this initial infiltration step, the water-saturated porous structure should enable autonomous liquid transport driven by capillary forces.

To keep the porous structure saturated with nutrient-rich liquid, the wicking rate needs to be comparable or higher than the evaporation rate. Our experiments on centimeter-scaled porous monoliths resulted in specific evaporation rates ranging from 97 to $125 \frac{g}{h\ m^2}$ and specific wicking rates ranging from 32 to $84 \frac{kg}{h\ m^2}$. These data provide useful guidelines for the design of brick elements, since it allows us to calculate the geometry that would ensure that the surface of the porous monolith remains wet (saturated condition). Based on our experimental data, the wicking rate will be higher than the evaporation rate if the ratio between exposed surface and cross-sectional area of the wall is equal or lower than approximately 500. For a brick element with a wicking cross-sectional area of 20x20cm$^2$, it should be possible to keep the porous structure saturated with liquid if the evaporation surface is kept below 20m$^2$.

Since these are reasonable dimensions for architectural walls and facades, our analysis suggests that our porous ceramics provide a suitable scaffold for the transport of nutrient-rich water for the growth of microorganisms in building elements.

**High salt concentrated (HSC) media**

The HSC media used for the cultivation of *Synechococcus sp* was prepared by mixing ASNIII and BG11 media in a 1 to 1 volume ratio.

The ASNIII medium was prepared using a standard recipe provided by the supplier (ATCC medium: ASN-III medium). 80% of deionized (MilliQ) water was first added to a beaker and heated to 50°C under stirring. The chemicals in Table I were added, the heater was turned off, and the solution left to stir over night at 600 rpm. Cyanocobalamin (Acros organics) was then added from a stock solution to reach a concentration of 10 µg mL$^{-1}$ and the remaining water was topped up. The media was autoclaved and let cool down below 60°C before the trace metal mix (Table II) was added in a trace metal mix: media volume ratio of 1:1000.

The BG11 medium was prepared by diluting the BG11 100x solution purchased from Merck with sterile deionized (MilliQ) water at a volume ratio of 1:100. The resulting solution was supplemented with 1:1000 trace metal mix (Table II).

## Supplementary figures

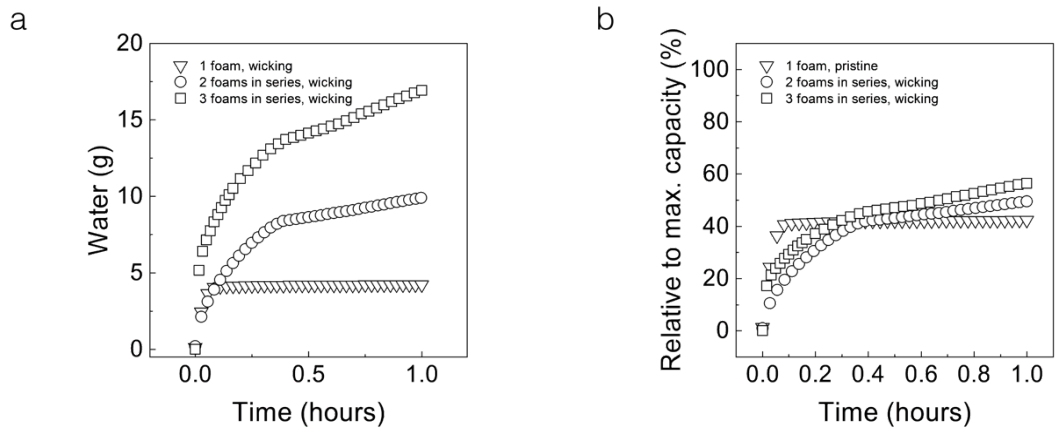

**Figure S1. Wicking behavior of 1, 2 and 3 porous monoliths in series.** Water uptake in absolute values (a) and relative to the maximum capacity of the pillar system (b). The maximum capacity was determined by infiltrating the structures under vacuum. By stacking multiple porous ceramic pillars on top of each other, more relative porosity can be filled up within the first hour. The meniscus forming at the interface between two pillars grows at a slower rate given its larger curvature. This additional wicking force is reflected in the larger slope for times longer than 0.4 hours in the case of the stacked systems, which is 0.05, 2.5, 5.5 mL/h in the case of 1, 2 and 3 pillars, respectively.

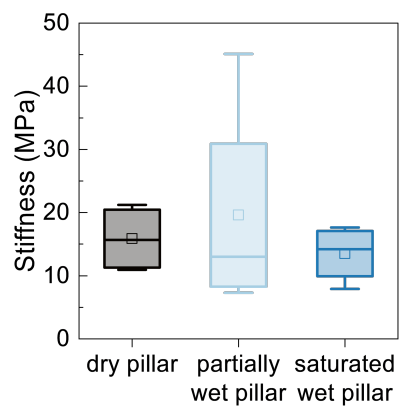

**Figure S2. Stiffness of cylindrical scaffolds in the dry, partially wet and fully wet states.**

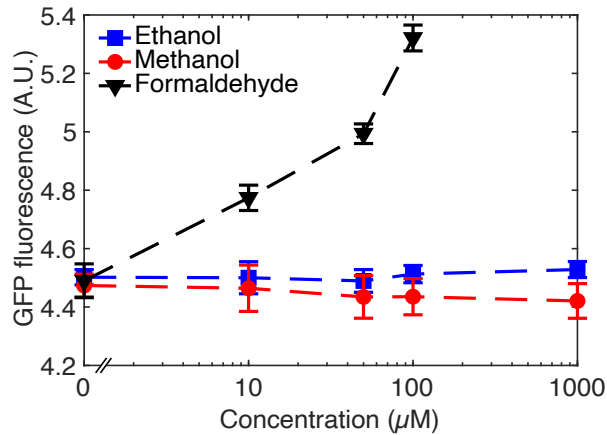

**Figure S3. Formaldehyde sensor cross-reactivity.** *E. coli* DH5a bearing plasmid pTR47m4-GFP were initially grown overnight in LB media with 100 μg mL$^{-1}$ carbenicillin at 37°C in a shaking incubator, and this starter culture was used to inoculate LB media containing varying concentrations of ethanol, methanol, or formaldehyde. Cultures were grown in 96-well plates at 30°C in shaking conditions. No GFP activation was observed with ethanol or methanol. No cell growth was observed at the highest formaldehyde concentration tested.

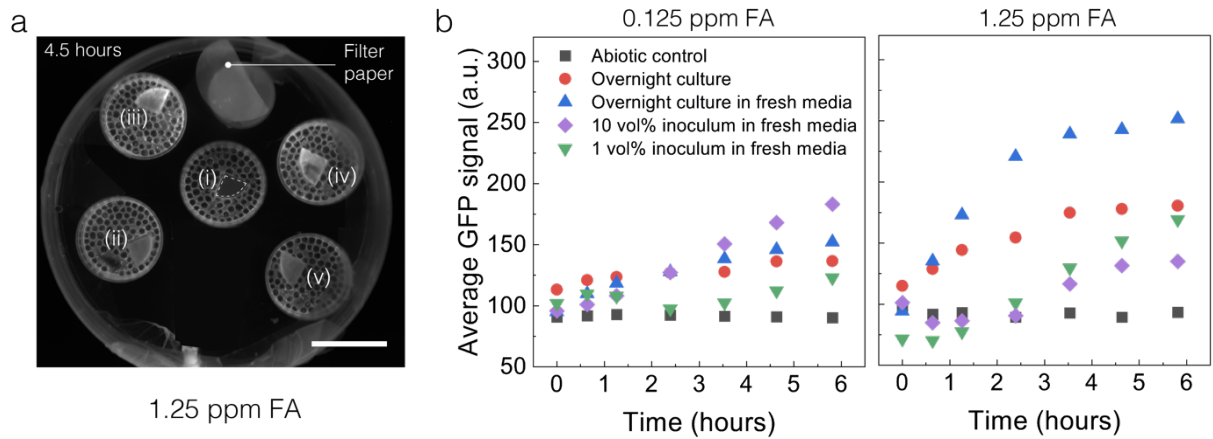

**Figure S4. Bacterial sensing of formaldehyde in the gas phase.** (a) Selected fluorescence image showing the experimental setup and the GFP signal of an abiotic control (i) and living porous ceramics infiltrated with an overnight culture (ii), an overnight culture redispersed in fresh medium (iii), 10 vol% inoculum in fresh media (iv) and 1 vol% inoculum in fresh media (v) of engineered *E. coli* exposed to 1.25 ppm of formaldehyde for 4.5 hours. Scale bar 5 cm. (b) GFP signal for living ceramics populated by different amounts of engineered *E. coli* exposed to 0.125 ppm (left) and 1.25 ppm (right) formaldehyde. The concentration of formaldehyde was estimated using Henry's law (see Figure S6).

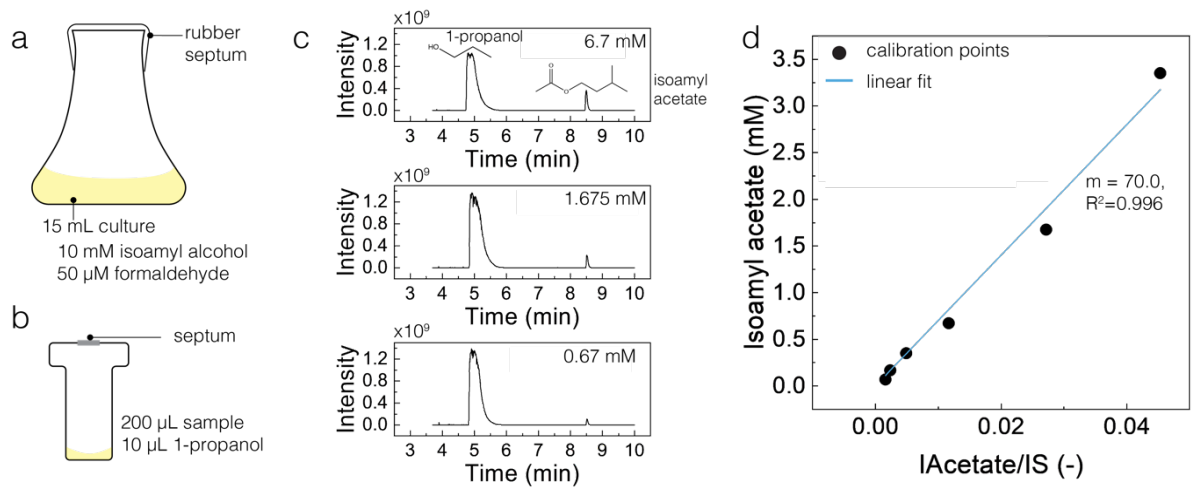

**Figure S5. Bacterial sensing of formaldehyde in the liquid phase.** (a) Schematic of the Erlenmeyer used to incubate the engineered *E. coli* DH5a bearing plasmid pFSKm4-ATF1 in the presence of formaldehyde for the conversion of isoamyl alcohol into isoamyl acetate. (b) Schematic of the headspace vial to measure the concentration of isoamyl acetate. (c) Selection of chromatographs for isoamyl acetate concentrations of 6.7, 1.675 and 0.67 mM. The first peak (RT = 4.9 min) is the internal standard (IS) 1-propanol and the smaller one (RT = 8.5 min) is isoamyl acetate (IAcetate). The retention time of each molecule was confirmed by the mass spectra. (d) The calibration curve was obtained by measuring the ratio of the peak integrals of IAcetate and IS for solutions containing known concentrations of isoamyl acetate in (c). The linear fit gives a slope (*m*) of 70.0 mM (R-square = 0.996).

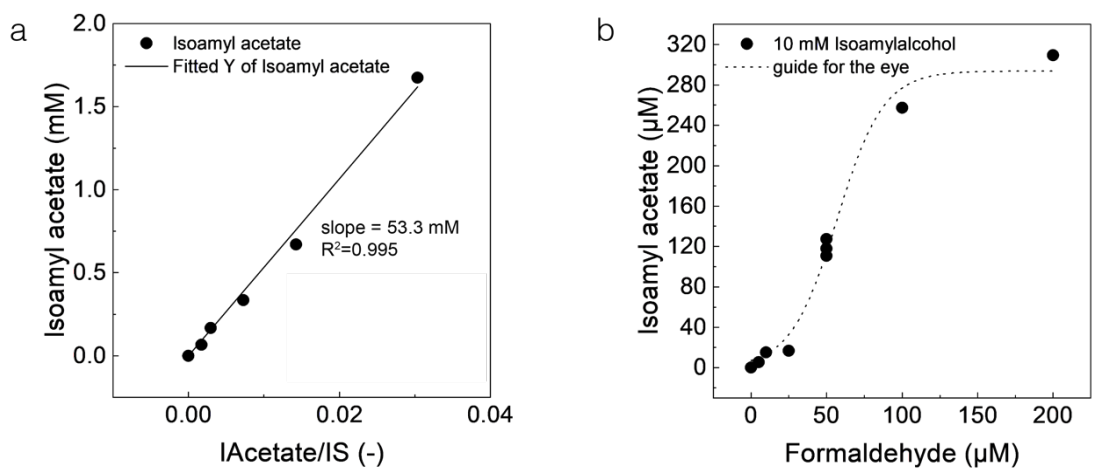

**Figure S6. Bacterial growth and sensitivity with formaldehyde in liquid.** *E. coli* DH5a bearing plasmid pFSKm4-ATF1 was initially grown overnight in LB media with 100 µg mL$^{-1}$ carbenicillin at 37°C in a shaking incubator, and this starter culture was used to inoculate LB media containing 10 mM isoamyl alcohol and varying concentrations of formaldehyde. Cultures were grown at 30°C in shaking conditions. (a) Calibration curve used to estimate the isoamyl acetate concentration. The internal standard 1-propanol was used. (b) Isoamyl acetate concentrations after 2 days growth in contact with formaldehyde. No isoamyl acetate activation was observed in absence of formaldehyde.

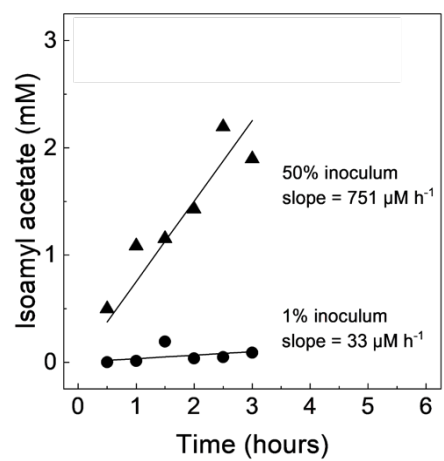

**Figure S7. Bacterial sensing of formaldehyde in liquid.** Conversion rates for 1% and 50% inoculum in the first 3 hours of the experiment shown in Figure 5c.

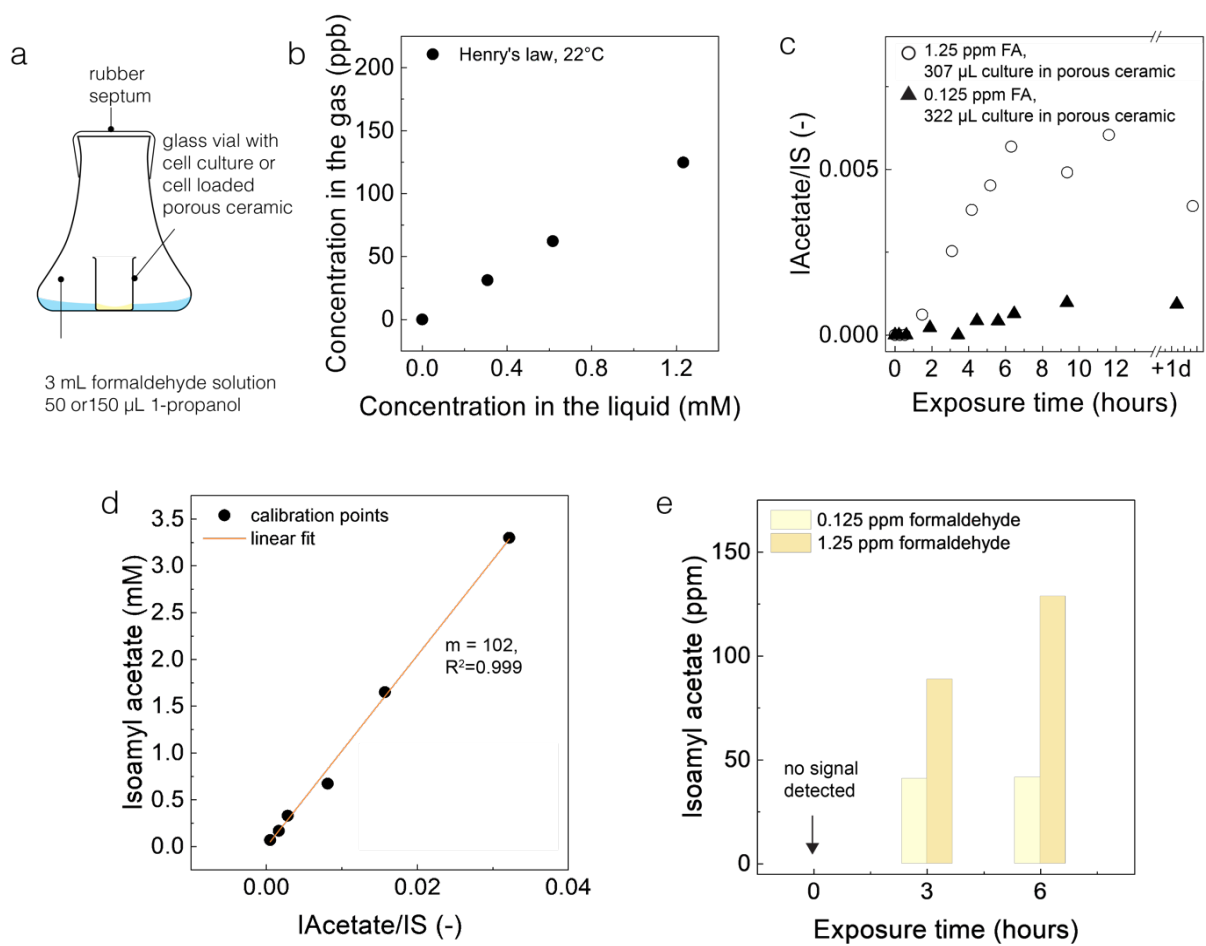

**Figure S8. Bacterial sensing of formaldehyde in the gas phase.** (a) Schematic of the Erlenmeyer from which headspace samples were taken. (b) Estimation of the concentrations of formaldehyde in the gas phase based on the partial pressure obtained from Henry's law: $p = H \cdot c$, where $p$ is the partial pressure of formaldehyde, $c$ is the concentration of formaldehyde in the liquid and the proportionality constant $H_{FA} = 0.1 \text{ atm cm}^3 \text{ mol}^{-1}$ is taken from values reported in literature for formaldehyde in water at 22°C.[4] (c) Isoamyl acetate (IAct) concentration relative to the internal standard (IS, 150 µL 1-propanol) for two porous ceramics containing the engineered *E. coli* DH5a bearing plasmid pFSKm4-ATF1 exposed to 125 and 1250 ppb formaldehyde (1.2 and 12 mM in the solution). (d) Calibration curve for the setup shown in (a) using 50 µL 1-propanol as internal standard. The linear fit gives a slope (*m*) of 102 mM (R-square = 0.999). (e) Isoamyl acetate concentration in the air after 0, 3 and 6 hours exposure time. The concentration in the air was estimated using Henry's law with the proportionality constant $H_{IAct} = 113.6 \text{ atm cm}^3 \text{ mol}^{-1}$ taken from literature values.[5]

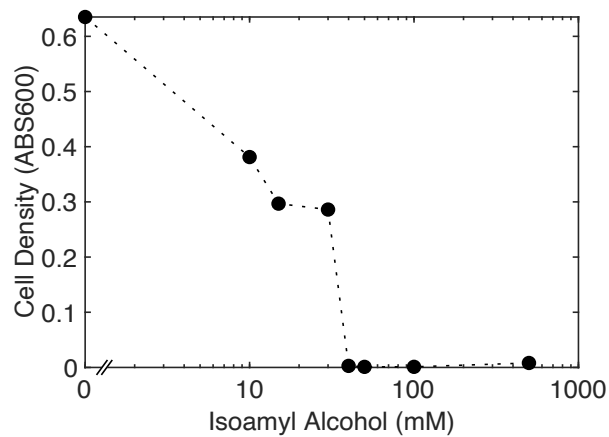

**Figure S9. Isoamyl alcohol toxicity for E. coli DH5a bearing plasmid pFSKm4-ATF1.** Bacteria were initially grown overnight in LB media with 100 µg mL$^{-1}$ carbenicillin at 37°C in a shaking incubator, and this starter culture was used to inoculate LB media containing a variable concentration of isoamyl alcohol, which was grown at 30°C under shaking conditions. No cell growth was detected above 30 mM.

# Supplementary tables

**Table I.** Ingredients for the ACSNIII media.

| Concentration (g L$^{-1}$) | Ingredient |
|---|---|
| 25 | Sodium chloride (VWR chemicals) |
| 0.95 | Magnesium chloride (Abcr) |
| 0.5 | Potassium chloride (VWR chemicals) |
| 0.02 | Di-potassium hydrogen phosphate trihydrate (p.a., Merck) |
| 3.5 | Magnesium sulfate heptahydrate (≥99.0%, Sigma Aldrich) |
| 0.5 | Calcium chloride dihydrate (VWR chemicals) |
| 0.003 | Citric acid (Sigma Aldrich) |
| 0.0005 | Ethylenediaminetetraacetic acid (EDTA) (≥99%, Roth AG) |
| 0.04 | Sodium carbonate (Merck) |
|  | Deionized (MilliQ) water |

**Table II.** Ingredients of the trace metal mix.

| Concentration (g L$^{-1}$) | Ingredient |
|---|---|
| 2.86 | Boric acid (≥99.5%, Sigma Aldrich) |
| 1.81 | Manganese(II) chloride tetrahydrate (99+%, Acros Organics) |
| 0.222 | Zinc sulfate heptahydrate (≥99.5%, Sigma Aldrich) |
| 0.390 | Sodium molybdate(VI) dihydrate (99+%, Acros Organics) |
| 0.079 | Cupric sulfate pentahydrate (≥99.0%, Fluka Chemie AG) |
| 0.049 | Cobalt(II) nitrate hexahydrate (≥98%, Sigma Aldrich) |
|  | Deionized (MilliQ) water |